\documentclass{ws-ijmpa}
\newcommand{\bea}{\begin{eqnarray}}
\newcommand{\eea}{\end{eqnarray}}
\newcommand{\be}{\begin{eqnarray}}
\newcommand{\ee}{\end{eqnarray}}
\newcommand{\bT}{{\bf b}_\perp}
\newcommand{\bp}{{\bf b}_\perp}
\begin{document}

\markboth{Matthias Burkardt}
{Hadron Tomography}

%
\catchline{}{}{}{}{}
%

\title{HADRON TOMOGRAPHY}

\author{\footnotesize MATTHIAS BURKARDT}

\address{Department of Physics, New Mexico State University\\
Las Cruces, NM 88011, USA
}

\maketitle

\pub{Received (Day Month Year)}{Revised (Day Month Year)}

\begin{abstract}
Generalized parton distributions can be used 
to obtain information about the dependence of parton distributions
on the impact parameter. Potential 
consequences for T-odd single-spin asymmetries are discussed.

\end{abstract}

\section{Introduction}	
Hadron form factors provide information about the Fourier transform
of the charge distribution within the hadron. Generalized
parton distributions (GPDs) provide a momentum decomposition
of the form factor w.r.t. the average 
momentum fraction $x = \frac{1}{2}\left(x_i+x_f\right)$ 
of the active quark
\bea
\int dx H_q(x,\xi,t) &=& F^q_1(t)
\quad \quad 
\int dx E_q(x,\xi,t) = F^q_2(t)
\eea
where $F^q_1(t)$ and $F^q_2(t)$
 are the Dirac and
Pauli formfactors, respectively.
$x_i$ and $x_f$ are the momentum fractions of the quark
before and after the momentum transfer.

The momentum direction of the active quark singles
out a direction in space and therefore it makes a difference
whether the momentum transfer is along that momentum or in a
different direction. Therefore GPDs not only depend on $x$
and the invariant momentum transfer $t$ but also on the
longitudinal momentum transfer through the variable
$2\xi=x_f-x_i$. 
Unlike form factors, in which the contribution from all
quark momenta are always summed up, GPDs thus tell us how much
each momentum contributes to the form factor at a given 
momentum transfer.

GPDs are the form factor of the
same operator whose forward matrix elements yield the usual
parton distribution functions (PDFs)
\bea
\int \frac{dx^-}{2\pi}e^{ix^-\bar{p}^+x}
\left\langle p^\prime \left|\bar{q}\left(-\frac{x^-}{2}\right)
\gamma^+ q\left(\frac{x^-}{2}\right)\right|p\right\rangle
&=& H(x,\xi,\Delta^2)
\bar{u}(p^\prime)\gamma^+ u(p)
\\
& &
+ E(x,\xi,\Delta^2)\bar{u}(p^\prime)
\frac{i\sigma^{+\nu}\Delta_\nu}{2M} u(p),
\nonumber\eea
This observation has formed the basis of the position space
interpretation of GPDs.

\section{Position Space Interpretation of GPDs}
Charge distributions in position space are usually measured
in the center of mass frame, i.e. measured relative to the
center of mass of the system. For impact parameter dependent
parton distributions, the analogous reference point is the
transverse center of momentum of all partons (quarks and gluons)
$
{\bf R}_\perp = \sum_{i=q,g} x_i {\bf r}_{\perp,i} ,
$
where $x_i$ is the momentum fraction carried by each parton and
${\bf r}_{\perp,i}$ is their $\perp$ position. 
This $\perp$ center of momentum behaves in many ways similar
to the nonrelativistic center of mass. For example, one can
form  eigenstates of ${\bf R}_\perp$
\bea
\left|p^+,{\bf R}_\perp={\bf 0}_\perp,\lambda
\right\rangle
\equiv {\cal N}\int d^2{\bf p}_\perp 
\left|p^+,{\bf p}_\perp,\lambda \right\rangle .
\eea
Impact parameter dependent PDFs are defined using the familiar
light-cone correlation function in such transversely localized
states
\bea
q(x,{\bf b}_\perp) \equiv\! 
\int \!\frac{dx^-\!\!\!}{4\pi\!} 
\left\langle p^+\!,{\bf R}_\perp={\bf 0}_\perp \right|
\bar{q}(-\frac{x^-\!\!}{2}\!,{\bf b}_\perp)
\gamma^+ q(\frac{x^-\!\!}{2}\!,{\bf b}_\perp)
\left|p^+\!,{\bf R}_\perp={\bf 0}_\perp\right\rangle 
e^{ixp^+x^-} ,
\eea
and similarly for the polarized distribution 
$\Delta q(x,{\bf b}_\perp)$ with an additional $\gamma_5$.

Impact parameter dependent PDFs are Fourier transforms of GPDs
for $\xi=0$
\bea
\label{master}
q(x, \bT) &=& \int \frac{d^2{\bf \Delta}_\perp}{(2\pi)^2 }
e^{i{\bf \Delta}_\perp\cdot \bT} H(x,0, - {\bf \Delta}_\perp^2)\\
\Delta q(x, \bT) &=& \int \frac{d^2{\bf \Delta}_\perp}{(2\pi)^2 }
e^{i{\bf \Delta}_\perp\cdot \bT} 
\tilde{H}(x,0, - {\bf \Delta}_\perp^2) \nonumber .
\eea
Due to a Galilean subgroup of transverse
boosts in the infinite momentum frame 
there are no relativistic corrections to Eq. (\ref{master}).
Furthermore, these impact parameter dependent parton distributions
have a probabilistic interpretation very similar (and with the
same limitations) as the usual PDFs\cite{mb:adl}. 

So far, only few experiments exist that help constrain GPDs and
therefore it is important to utilize theretical constraints when
parameterizing these functions. One such constraint arises directly
from the fact that the reference point for
impact parameter dependent PDFs is the $\perp$ center of momentum.
In the limit of $x\rightarrow 1$ the active quark becomes the
center of momentum and therefore $\bT$ can never be large for large 
$x$. For impact parameter dependent parton distributions this 
implies that their $\perp$ width should go to zero for 
$x\rightarrow 1$. For decreasing $x$ the $\perp$ width is expected
to gradually increase. Although the width in the valence region 
should still be relatively compact, its size should increase further once $x$ is small enough for the pion cloud to contribute \cite{weiss}.
In momentum space, this implies that the $t$-dependence of GPDs
should decrease with increasing $x$. This prediction is consistent 
with recent lattice results, which showed that higher moments of GPDs
have less $t$ dependence than lower moments \cite{lattice1}

\section{Transversely Polarized Target}
For a transversely polarized target, the impact parameter dependent
parton distributions are no longer axially symmetric. The deviations
from axial symmetry are described by $E(x,0,t)$. For example, 
the unpolarized quark distribution $q_X(x,\bT)$ for
a target that is polarized in the $+\hat{x}$ direction reads
\cite{IJMPA}
\be
q_X(x,\!{\bf b_\perp}) = q(x,\!{\bf b_\perp})
-
\frac{1}{2M}\frac{\partial}{\partial b_y}\!
\int \!\!\frac{d^2{\bf \Delta}_\perp }{(2\pi)^2} 
E(x,0,\!-{\bf \Delta}_\perp^2)
e^{-i{\bf b}_\perp\cdot{\bf \Delta}_\perp}.
\ee
Here $q(x,\!{\bf b_\perp})$ is the impact parameter 
dependent PDF in the unpolarized case (\ref{master}).
The origin of this distortion is that the virtual
photon in DIS couples more strongly to quarks that move
towards it than quarks that move away from it (hence the
$\gamma^+$ in the quark correlation function relevant for DIS).
If  the orbital motion of the
quarks and the spin of the target are correlated
then quarks are more likely
to move towards the virtual photon on one side of the target than 
the other and therefore the distribution of quarks in impact
parameter space appears deformed towards one side. The details
of this deformation for each quark flavor
are described by $E_q(x,0,t)$, which is not known yet.
However, sign and overall scale can be estimated by considering
the mean displacement of flavor $q$
($\perp$ flavor dipole moment) 
\be
\label{dipole}
d^q_y \equiv \int\!\! dx\!\int \!\!d^2\bp
q(x,\bp) b_y
=\frac{1}{2M} 
\int\!\! dx E_q(x,0,0) = \frac{\kappa_{q}^p}{2M} .
\ee
\begin{figure}
\unitlength1.cm
\begin{picture}(10,3.5)(.5,10.3)
\includegraphics{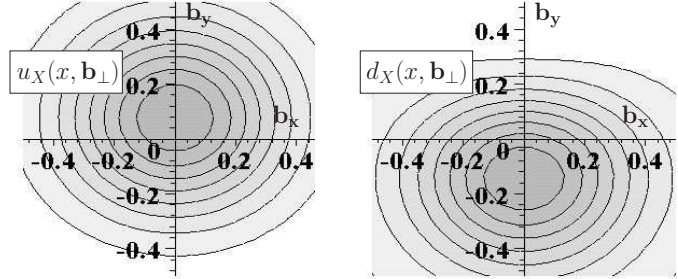}
\end{picture}
\caption{Distribution of the $j^+$ density for
$u$ and $d$ quarks in the
$\perp$ plane ($x_{Bj}=0.3$ is fixed) for a nucleon that is 
polarized in the $x$ direction in the model from
Ref. 
For other values of $x$ the distortion looks similar.
More recent fits yield similar results.
}
\label{fig:distort}
\end{figure}  
The $\kappa_q={\cal O}(1-2)$ 
are the anomalous magnetic moment contribution
from each quark flavor to the anomalous magnetic moment of the
nucleon, i.e.
$F_2(0) = \frac{2}{3} \kappa_u - \frac{1}{3}\kappa_d-\frac{1}{3}\kappa_s...$.
This yields $\left|d^q_y\right| ={\cal O}(0.2 fm)$, where
$u$ and $d$ quarks have opposite signs. This is a sizeable
effect as is illustrated in Fig. (\ref{fig:distort}).

\begin{figure}
\unitlength1.cm
\begin{picture}(10,2)(3.,19.2)
\includegraphics{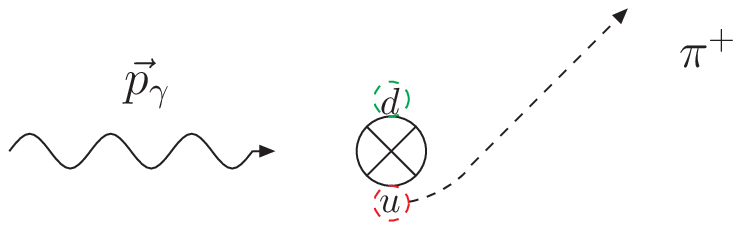}
\end{picture}
\caption{The transverse distortion of the parton cloud for a proton
that is polarized into the plane, in combination with attractive
FSI, gives rise to a Sivers effect for $u$ ($d$) quarks with a
$\perp$ momentum that is on the average up (down).}
\label{fig:deflect}
\end{figure}

This deformation provides a very physical source for single-spin
asymmetries (SSA) in semi-inclusive DIS. For an (on average)
attractive final 
state interaction, the position
space deformation into the $+\hat{y}$ direction translates
into a momentum space asymmetry for the ejected quark that prefers
the $-\hat{y}$ direction and {\it vice versa} 
(Fig. \ref{fig:deflect})
Since the sign of the position space distortion is governed by
the sign of the anomalous magnetic moment contribution $\kappa_{q/P}$
from each
quark flavor, this implies that the sign of the SSA is correlated
to the sign of $\kappa_{q/P}$. Following the Trento convention 
\cite{trento}, this yields a negative Sivers
function $f_{1T}^{\perp u}$ for $u$ quarks in the proton, while
$f_{1T}^{\perp d}>0$.
For the neutron those signs are
reversed. These predictions are consistent with recent HERMES
data \cite{HERMES}.

\section{Chirally Odd GPDs}
The distribution of transversely polarized quarks in impact 
parameter space is descibed by the Fourier transform of
chirally odd GPDs \cite{dh}. For example, for an unpolarized
target the distribution of quarks with transversity $s^i$
reads
\be
q_i(x,\bT)= -\frac{s^i\varepsilon^{ij}}{2M} 
\frac{\partial}{2b_j} \int \frac{d^2{\bf \Delta}_\perp}{(2\pi)^2}
\left[ 2{\tilde H}_T(x,0,-{\bf \Delta}_\perp^2)
+ E_T(x,0,-{\bf \Delta}_\perp^2)\right]
e^{-i\bT\cdot{\bf \Delta}_\perp}.
\ee
A consequence of this result is an analog of Ji's
angular momentum sum rule, where the nucleon spin 
is replaced
by the quark spin: the angular momentum $J^i_q$ carried by quarks
with transversity $s^j$ in an unpolarized target reads
\cite{mb:odd}
\be
\left\langle J_q^i(s^j)\right\rangle = \frac{\delta^{ij}}{4}
\int dx\, x \left[H_T(x,0,0)+2 \tilde{H}^q_T(x,0,0) + E^q_T(x,0,0)\right]. 
\label{Jodd}
\ee
The relation between Eq. (\ref{Jodd}) and Ji's sum rule \cite{ji}
is similar to the relation between the Sivers function and the
Boer-Mulders function $h_1^{\perp q}$ \cite{BM} in that the nucleon spin is
replaced by the quark spin. This observation suggests extending
the physical mechanism for the Sivers effect (Fig. \ref{fig:deflect})
to the Boer-Mulders effect, yielding opposite signs for
$2 \tilde{H}^q_T(x,0,0) + E^q_T(x,0,0)$ and $h_1^{\perp q}$.

\section*{Acknowledgments}

This work has been supported
by the DOE (grant number DE-FG03-95ER40965).




\begin{thebibliography}{0}

\bibitem{mb:1st} M. Burkardt, Phys.\ Rev.\  {\bf D62}, 071503 (2000);
Erratum-ibid. {\bf D66}, 119903 (2002).

\bibitem{mb:adl} M. Burkardt, hep-ph/0105324;
P.V. Pobilitsa, Phys.\ Rev.\ {\bf D70}, 034004 (2004). 

\bibitem{weiss} M. Strikman and C. Weiss, Phys.\ Rev.\ {\bf D69},
054012 (2004).

\bibitem{lattice1} 
Ph. H\"agler et al., Phys. Rev. Lett. {\bf 93}, 112001 (2004).

\bibitem{IJMPA} M. Burkardt, 
Int. J. Mod. Phys. {\bf A18}, 173 (2003).

\bibitem{ji+} X.Ji, Phys. Rev. Lett. {\bf 78}, 610 (2003).

\bibitem{trento} A.Bacchetta et al., Phys.\ Rev.\ {\bf D70}, 
117504 (2004).

\bibitem{HERMES} HERMES collaboration, Phys.\ Rev.\
Lett. {\bf 94}, 012002 (2005).

\bibitem{dh} M. Diehl and Ph. H\"agler, hep-ph/0504175.

\bibitem{mb:odd} M. Burkardt, hep-ph/0505189.

\bibitem{ji} X. Ji, Phys.\ Rev.\ Lett.\ {\bf 78}, 610 (1997).

\bibitem{BM} D. Boer and P.J. Mulders, Phys.\ Rev.\ {\bf D57}, 5780
(1998).

\end{thebibliography}
\end{document}